\documentstyle[12pt]{article}

\newcommand{\beq}{\begin{equation}}
\newcommand{\eeq}{\end{equation}}
\newcommand{\beqa}{\begin{eqnarray}}
\newcommand{\eeqa}{\end{eqnarray}}

\begin{document}
\begin{center}
{\bf \large Exclusive Reactions in QCD\footnote{lecture given at the Les Houches
Summer School:  Trends in Nuclear Physics Hundred Years Later, July 30 to
August 30 1996,
 to be published in the Proceedings, H. Niffenecker ed.( Elsevier)}}
\vspace{1cm}

Bernard PIRE
\vspace{1cm}

Centre de Physique Th\'eorique\footnote{Unit\'e propre 14 du Centre
National
de la Recherche Scientifique.}
         Ecole Polytechnique
          F91128 Palaiseau

\end{center}

\vspace{1cm}

\section{Introduction}

We review the theory of hard exclusive scattering in Quantum Chromodynamics.
 After recalling the classical counting rules which describe the leading scale
dependence of form factors and exclusive cross-sections at fixed angle, the
 pedagogical example of the
pion form factor is developped in some detail in order to show explicitely what
factorization means in the QCD framework. The picture generalizes to many
hard reactions which are at the heart of the ELFE project~\cite{ELFE}. We
briefly
present the concepts of color transparency.

\section{Space-time picture and Counting Rules }

We consider here {\it exclusive processes }, that is
interactions  resulting in a  final state where all
particles are identified. Using a
perturbative expansion to study these reactions may {\it a priori}
be foreseen if a large momentum transfer appears: this is what is called a
{\it hard} reaction.

Before going to a QCD calculation, it is very instructive to develop first a
space-time picture of these reactions.

The simplest exclusive quantity is the {\it pion form factor}, as measured in
 the process $e^-\pi^+\rightarrow e^-\pi^+$. It measures the ability of the
 pion to stay itself when being collided by an electron. It is thus a quantity
much sensitive to  confinement mechanisms. The physics deals with the
restauration of the meson integrity after the violent shock of a
 high-energy electron with one of the quarks. At the limit
$Q^2=0$, the meson  structure is not resolved, and $F_{\pi}(0)=1$.

Let us first derive in a heuristic way the $Q^2$ dependence of the pion
form factor
 by a careful examination of the way this process takes place.
In its rest frame, the pion is represented as a
collection of partons,  quarks and gluons, approximately
uniformously spreaded in a  sphere of  radius $R$ (typically the
pion charge radius, around $0.5\,$fm). In the reaction center of mass frame,
where the final electron emerges
 at an angle of 180$^{\circ}$ with respect to the initial electron,
the longitudinal  dimension is Lorentz-contracted to $R/\gamma$ with
$\gamma=Q/2M$. The transverse dimensions are on the other hand not
affected by Lorentz-contraction. At time 0, the electron hits one of the
 partons, the so-called {\it active} parton, and both change directions.
For the whole process to be elastic, {\it all} other partons
must be alerted before the moment $t\approx1/Q$ to
form the emerging pion  (also contracted in this frame).
The motion of the active parton after the collision is
$z(t)=-t,\ x(t),y(t)=0$ whereas the motion of a  {\it spectator} parton is
$z(t)=t+z_0,\ x(t)=x_0,\ y(t)=y_0$ (one has $-1/Q< z_0< 1/Q$ and
$-R\le x_0,y_0\le R$). Between the moments 0 and $1/Q$ a spectator parton
can receive and respond to a physical signal emitted by the active parton
at time 0
only if the interval $\Delta=t^2-(t+z_0)^2-x_0^2-y_0^2$ is positive, that is if
the spectator is at a distance $\sqrt{x_0^2+y_0^2}<1/Q$ in the transverse
plane. One thus counts the probability to find spectator partons in a
transverse disc of  radius $1/Q$, in the initial as well as in the  final
state.
One gets\cite{Bro73}
\begin{equation}
F_{\pi}^2\propto\left({\pi Q^{-2}\over \pi R^2_{\pi}}\right)
^{n_{in}-1+n_{out}-1}.
\end{equation}
Since a pion contains at least a valence quark and antiquark, we get a
minimal contribution scaling like $1/Q^2$. Adding for instance one
 gluon to the  valence in the initial state, without changing
the final state, yields a contribution scaling like $1/Q^3$ \dots\ These
contributions diminish relatively to the valence state contribution
 as energy increases.

This most important feature of the study of form factors at large transfer
may be generalized to other exclusive reactions \footnote{with the important
exception of hadron hadron exclusive reactions for which a separate study has
been developped\cite{Lan,Mue81,BS89} with the important result of valence
selection but with
different transverse size and scaling behaviours}: {\it when the interaction is
at short distance, the valence  contributes in a  dominant way in terms of
scaling law}. Moreover, and this will be crucial for the phenomenon of
{\it color transparency}, the hadron configurations which contribute have
small ($O(1/Q)$) transverse sizes.

Let us summarize: asymptotically, one predicts for the energy dependence of
pion and nucleon form factors
\footnote{In the proton case, there are two form factors and the reasonning
developped here does not allow to distinguish them.
In fact, if one separates the form  factors with respect to their  degree
of helicity conservation,
one shows that the above counting rule applies only for helicity
conserving processes (and thus for the magnetic form factor $G_M$), but that
an additional power suppression affects $F_2$.}, a power-law fall off:
\begin{equation}
F_{\pi}(Q^2)\propto {1\over Q^2}~~~~~~~~~~~~~~
F_N(Q^2)\propto {1\over Q^4}.
\end{equation}
and for exclusive reactions differential cross sections at fixed angle {\it
i.e.}
 at large $s$, $-t$ and $-u$~\cite{Bro73}:

\begin{equation}
{d\sigma\over dt} = {1\over  s^{N-2}} f({t\over s}),
\end{equation}
where N is the total number of elementary fields participating in the reaction
One thus gets that the transition between {\it  valence} states dominates hard
amplitudes, {\it i.e.} we get a leading contribution with N=8 for Compton
scattering,
with N=9 for meson photoproduction....
 The QCD analysis presented in section 3 will
strengthen the argument presented here and develop a consistent way of
calculating the leading contribution. It will however be important to
phenomenologically
verify that the scaling laws, and thus the dominance of valence states, apply
 at accessible energies, and this for each physical process under study.

\section{Calculating a hard exclusive amplitude: the example of the  pion
form factor }

One now wants to really calculate from QCD a hard amplitude~\cite{Far79};
we will here detail
the procedure in one of the simplest example, namely the electromagnetic  pion
 form factor at large spacelike transfer $Q^2$.
 This leads us to precise first the hadron wave function and the
Born hard amplitudes, then the radiative corrections to see if a sensible
picture emerges where a non perturbative object sensitive to confinement
dynamics
factorizes from a hard scattering amplitude controlled by a perturbative
expansion
in $\alpha_s(Q^2)~$ which is renormalization group improved. This
factorization which is
crucial for a consistent understanding of future experimental data turns
out to be valid
for many hard exclusive reactions\footnote{
indeed factorization also holds for hadron-hadron reactions\cite{BS89}}. It
 may be pictorially described as in Fig.1.

\vglue5cm
\centerline{\small Figure 1: Factorization of a hard exclusive process :
$X*T_H*X'$}

\subsection{Description of the pion}

Let us specify the kinematics. In the Breit frame
the momenta are written as:
\begin{equation}
q=\pmatrix{0\cr0\cr0\cr Q\cr},\
p=\pmatrix{Q/2\cr0\cr0\cr -Q/2\cr},\
p'=\pmatrix{Q/2\cr0\cr0\cr Q/2\cr};
\end{equation}
where the pion mass is neglected in front of $Q$.

To describe the  pion  in its  valence state, one introduces  the
Bethe-Salpeter~(BS) amplitude~\cite{Sal51}
\begin{equation}
\langle0|T\left(q_{u\alpha i}(y)~P_{ij}(y,0)~\bar{q}_{d\beta
j}(0)\right)|\pi^+(p)\rangle,
\end{equation}
where $u$ and $\bar{d}$ are the flavours of  the valence quarks of $\pi^+$,
$\alpha$ and $\beta$ are Dirac indices and $i$, $j$ are color
indices. The $P_{ij}$ operator is necessary to have an amplitude invariant under
local gauge transformations; when $q(y)$ transforms to $U(y)~q(y)$, $P(y,0)$
transforms to $ U^{-1}(y)~P(y,0)~U(0) $, compensating the quark and antiquark
 variations. The BS amplitude is the relativistic generalisation of the
 Schr\"odinger wave function  describing the bound state
of a quark antiquark pair. One may interpret it as the
 probability amplitude of finding in a $\pi^+$ a $u$ quark at
point $y$ and a  $\bar{d}$ antiquark at the origin.

One often prefers to work in momentum space and defines the Fourier
 transform of the BS amplitude as
\begin{equation}
\int d^4ye^{ik.y}\langle0|T\left(q_{u\alpha i}(y)~P_{ij}(y,0)~\bar{q}_{d\beta
j}(0)\right)|\pi^+(p)\rangle =X_{\alpha\beta}(k,p-k)
\end{equation}
where $k$ is the quark momentum and, by momentum conservation,
$p-k$ is the antiquark momentum.

To discuss the properties of this amplitude, it is convenient to
introduce light-cone coordinates defined as\footnote{The  scalar product of two
$4-$vectors is then
$A.B=A^+B^- +A^-B^+ -{\bf A}_{\bot}.{\bf B}_{\bot}$.
}:
\begin{equation}
\left\{\matrix{
k^+={1\over \sqrt{2}}(k^0-k^3)\cr
k^-={1\over \sqrt{2}}(k^0+k^3)\cr}\right\}
\end{equation}

We thus have (listing $p=[p^+,p^-,p^1,p^2]$)
\begin{equation}
p=[Q/\sqrt{2},0,0,0],\ p'=[0,Q/\sqrt{2},0,0],
\end{equation}
and we parametrize the internal momenta as
$k=[xQ/\sqrt{2},k^-,{\bf k}_{\bot}]$,
where $x$ is the light-cone fraction carried by the  quark inside the pion.
The antiquark then carries the fraction $1-x=\bar{x}$. The final  pion is
treated similarly, with $+$ and $-$ components exchanged:
$k'=[k'^+,x'Q/\sqrt{2},{\bf k'}_{\bot}]$.

In terms of these  variables, the $k^{\mu}$ regions  favored by the
amplitude $X(k,p-k)$ are simply written as:
\begin{equation}
{\bf k}_{\bot}^2< M^2,\hskip 1cm |k^-|< M^2/Q.
\end{equation}

\subsection{The hard scattering at the Born level}

The matrix element of  Figure~1 is written as the convolution
\begin{equation}
\int {d^4k\over (2\pi)^4}{d^4k'\over (2\pi)^4}\,
X(k)\,T_H^{\mu}(k,k')\,X^{\dag}(k').
\end{equation}
At the lowest order in the QCD coupling constant, $g$, one finds 4
Feynman diagrams. One is drawn on   Figure~2 and the 3 others are easily deduced
by attaching successively the photon to the  points 2, 3 and 4.

Let us first evaluate the gluon squared momentum, which is in Feynman gauge,
the denominator of the gluon propagator. We have
\begin{equation}
\label{virtualite}
\matrix{
(p'-k'-p+k)^2=&-\bar{x}\bar{x}'Q^2
&-\sqrt{2}Q(k^-\bar{x}'+k'^-\bar{x})&-2k^-k'^+
&-({\bf k}_{\bot}-{\bf k}'_{\bot})^2\cr
\noalign{\medskip}
&O(Q^2)&O(M^2)&O({\displaystyle M^4\over\displaystyle Q^2})&O(M^2)\cr
}\nonumber
\end{equation}
where typical orders of magnitude indicated refer to the momentum regions
favored by the amplitudes $X(k)$ and $X(k')$. Restricting to leading terms in
$Q$, we may forget terms of order $M^2$. So, in particular, we write
\begin{equation}
\label{virtualiteapprox}
(p'-k'-p+k)^2\approx-\bar{x}\bar{x}'{\displaystyle Q^2\over\displaystyle 2}.
\end{equation}

\vglue5cm
{\center \small Figure 2: Born Graph for the pion form  factor; the 3 other
graphs
are deduced by attaching the photon to the  points 2, 3 and 4. Propagators
joining   Bethe-Salpeter amplitudes to the vertices are
absorbed, by definition, in these amplitudes.}

\vspace{0.2in}
The same analysis may be repeated for the other quantities present
in the hard amplitude  $T_H^{\mu}$, leading to
\begin{equation}
T_H^{\mu}(k,k')\approx
T_H^{\mu}\left(x{Q\over\sqrt{2}},x'{Q\over\sqrt{2}}\right).
\end{equation}
We may then express the convolution of equation~(3.7) under the
form
\begin{equation}
\int dxdx'\,
\left({Q\over2\sqrt{2}\pi}\int {dk^-d{\bf k}_{\bot}\over (2\pi)^3}X(k)\right)
T_H^{\mu}(x,x')
\left({Q\over2\sqrt{2}\pi}
\int {dk'^+d{\bf k'}_{\bot}\over (2\pi)^3}X^{\dag}(k')\right),
\end{equation}
and the object needed to describe the pion in this reaction is in fact much
simpler than the amplitude $X$ since one may integrate over three components
of the internal momentum.

A first simplification comes from the integration over $k^-$ (for the
 outgoing pion over the $k'^+$).
In  terms of the conjugated variable $y^+$, this means that one only needs the
 Bethe-Salpeter amplitude at $y^+=0$, which is called the light cone wave
 function, usually noted as  $\psi(x,{\bf k}_{\bot})$~\cite{Bro}. A
useful property of this wave function is that the support in the  $x$
fraction is limited, as $0\le x\le 1$.

The Dirac structure of the amplitude $X(k)$ integrated over $k^-$ and
${\bf k}_{\bot}$ is easy to extract~\cite{Gou} and one finds
\begin{equation}
M_{\alpha\beta}(x,p)=
{1\over4}\gamma^5p\slash\,\phi(x)\big|_{\alpha\beta}.
\end{equation}
This Dirac structure corresponds to the combination of spinors
($\uparrow$ and $\downarrow$ denote respectively the helicity states $+$
and $-$)
\begin{equation}
{1\over4}\gamma^5\,p\slash|_{\alpha\beta}=
{1\over2\sqrt{2x\bar{x}}}{1\over\sqrt{2}}\left(
u_{\alpha}(x{\bf p},\uparrow)\,\bar{v}(\bar{x}{\bf p},\downarrow)
-u_{\alpha}(x{\bf p},\downarrow)\,\bar{v}(\bar{x}{\bf p},\uparrow)\right),
\end{equation}
{\it i.e.} one recovers the pion spin wave function in the quark  model
${1\over\sqrt{2}}\left(|\uparrow\downarrow\rangle
-|\downarrow\uparrow\rangle\right)$.

The function $\phi(x)$ is called the {\it distribution amplitude}; it
``measures'' how the  pion momentum is  distributed between the valence
quark and
antiquark when their transverse separation vanishes. {\it This is the
 non perturbative amplitude  connecting long distance physics
of strong interaction to short distance hard processes}.

Let us now precise a little bit the color algebra involved here. A useful
way to
simplify this matter is to choose for a pion of momentum along the $+$
direction,
axial gauges with axis along the  $-$ direction ( fixing $A^+=0$). In
these gauges, one has $P_{ij}(y,0)=\delta_{ij}$ and the color component for the
quark-antiquark pair is simply $\delta_{ij}/3$.
This fact partly explains the interest of light-cone gauges in the study
of hard processes. For another gauge choice, an explicit form of $P_{ij}(y,0)$
is necessary, but $P_{ij}(y,0)$
may be perturbatively analyzed and gauge invariance preserved order by
order in the perturbative expansion. At zeroth order, one has
\begin{equation}
P_{ij}(y,0)=\delta_{ij}+O(g).
\end{equation}

We are now able to calculate the graph of Figure~2 with a new Feynman rule
for the pion
\begin{equation}
{1\over 3}\delta_{ij}{1\over4}\gamma^5\,p\slash|_{\alpha\beta}\,\phi(x),
\end{equation}
and a loop integral $\int_0^1dx$. The amplitude of the  process may thus be
written as
\begin{equation}
\int_0^1dx\int_0^1dx'\phi(x)\langle T_H^{\mu}(x,x')\rangle\phi^*(x')
\end{equation}
where the hard process is evaluated on the spin and color components written
above. Color algebra leads to the trace
\begin{equation}
{1\over3}\delta_{ij}T^a_{jk}{1\over 3}{\delta_{kl}\over3}T^a_{li}
={C_F\over3}={4\over 9},
\end{equation}
and the amplitude  neglecting quark masses is
\begin{eqnarray}
\int_0^1dx\int_0^1dx'
(-){C_F\over3}Tr\left\{e_u\gamma^{\mu}{1\over4}\gamma^5p\slash\,
g\gamma^{\alpha}{1\over4}\gamma^5p\slash'\,g\gamma^{\beta}
{p'-\bar{x}p\over -\bar{x}Q^2}\right\}
\nonumber\\
{-\eta_{\alpha\beta}\over -\bar{x}\bar{x}'Q^2}\phi(x)\phi^*(x')
 = e_up^{\mu}{C_Fg^2\over6Q^2}
\left|\int_0^1dx{\phi(x)\over\bar{x}}\right|^2.
\end{eqnarray}

The graph with the photon  attached to  point 2 leads to the same
expression replacing $p^{\mu}$ by $p'^{\mu}$. The two other graphs
areidentical to the two first ones after exchanging $e_u\leftrightarrow-e_d$
and $\bar{x}\leftrightarrow x$ in the integrand denominator.
Charge  conjugation invariance and  isospin symmetry lead to the
relation $\phi(x)=\phi(\bar{x})$, so that one can factorize the
term $(e_u-e_d)(p+p')^{\mu}$ expected in Eq.~(8) and
isolate the form factor expression
\begin{equation}
F_{\pi}(Q^2)=
{C_Fg^2\over6Q^2}\left|\int_0^1dx{\phi(x)\over\bar{x}}\right|^2.
\end{equation}

Let us stress that we recover the scaling law in  $Q^{-2}$ predicted
by the counting rules.
\smallskip

The pion lifetime fixes a constraint on the valence wave function of the
pion since
 one may isolate the weak transition at the quark
level under the form of the matrix element of the electroweak current. One gets
\begin{equation}
\langle0|\bar{q}_d(0)\gamma^{\mu}(1-\gamma^5)q_u(0)|\pi^+(p)\rangle
= \sqrt{2}f_{\pi}p^{\mu},
\end{equation}
where the decay constant, $f_{\pi}$, is approximately equal to 92 MeV. This
leads to

\begin{equation}
\int_0^1dx\,\phi(x)=\sqrt{2}f_{\pi},
\end{equation}
which fixes the normalization of the distribution amplitude. Let us stress
that there is no
such normalization condition for the proton distribution amplitude, unless
one measures
proton decay...

\subsection{Radiative corrections}

It is important, when calculating a quantity in any field theory, and in
particular in  perturbative QCD, to keep track of  radiative corrections
and control them so that the picture obtained at lowest order survives their
inclusion.
The ultraviolet regimes does not a priori cause much problem since the
 theory is known to be  renormalizable. In fact, the subtractions
to be taken into account  are automatically taken care of when
 correctly treating quark and gluon propagators on the one hand,
and the running coupling constant on the other hand.

The infrared regions in the loop calculations must be very carefully
scrutinized.
In the specific process studied here, one finds in a  $n$ loops diagram
corrections of order
\begin{equation}
{\alpha_S(Q^2)\over Q^2}\left[\alpha_S(Q^2)\ln {Q^2\over M^2}\right]^n,
\end{equation}
which, since $\alpha_S(Q^2)\propto(\ln Q^2/\Lambda^2)^{-1}$ is of
the same order as the tree level process! One has to resum these large
logarithms in the distribution amplitude to recover the predictibility
of the  formalism. This is {\it factorization} since then the process may
be written as the convolution illustrated by Figure~2:
\begin{equation}
F_{\pi}=\phi*T*\phi^*
\end{equation}
where:

-- $T$ is a hard amplitude that one can evaluate within
 perturbative  QCD; namely, higher order corrections to $T$ are of order
 $\alpha_S^n(Q)$, and thus sufficiently small at sufficiently
large transfer;

-- all large  logarithms are absorbed in $\phi$; the
distribution $\phi$, which represents the wave function evolves
with the scale $Q$ characteristic of the virtual photon probe.
This stays an essentially non perturbative quantity expressing the way
confined valence quarks share the hadron momentum when they interact at
small distance in an exclusive process.

Let us now examine how leading logarithms are resummed in the distribution
 $\phi$~\cite{Field}. It turns out that it is most interesting to
choose to work in a gauge which is different from the Feynman gauge, namely
an axial gauge, with axis $n^{\mu}$,  fixing the  condition on gluon fields
 $A^{\mu}$ as: $n.A=0$. The leading  corrections have then the
 form illustrated on Figure~3.

One may show that the graph summation  yields
\begin{eqnarray}
\phi(x,Q)&=&\,\phi_0(x)
+\kappa\int_0^1du\,V_{q\bar{q}\rightarrow
q\bar{q}}(u,x)\,\phi_0(u)\nonumber\\
&+&{\kappa^2\over2!}\int_0^1duV_{q\bar{q}\rightarrow q\bar{q}}(u,x)
\int_0^1du'V_{q\bar{q}\rightarrow q\bar{q}}(u',u)\,\phi_0(u')+\ldots
\end{eqnarray}
where
$\kappa$ contains large collinear logarithms under the form
\begin{equation}
\kappa={4\over
\left(11-{2\over3}n_f\right)}\ln{\alpha_S(\mu^2)\over\alpha_S(Q^2)};
\end{equation}
and $V_{q\bar{q}\rightarrow q\bar{q}}$ is a characteristic kernel describing
the splitting of the
 valence distribution of the pion
\begin{equation}
V_{q\bar{q}\rightarrow q\bar{q}}(u,x)={2\over3}\left\{
{\bar{x}\over\bar{u}}\left(1+{1\over u-x}\right)_+\theta(u-x)+
{x\over u}\left(1+{1\over x-u}\right)_+\theta(x-u)\right\},
\end{equation}
The $()_+$ distribution comes from the compensation of infrared divergences
 (here in the limit $u\rightarrow x$)  between
 graphs b and c of  Figure~3. This is a consequence
of the colour neutrality of a hadron.

\vglue5cm
\centerline{\small Figure 3: Leading corrections in axial gauge}

\vspace{0.2in}
The equation on $\phi$  may be rewritten
under the integro-differential form
\begin{equation}
\left({\partial\phi\over\partial\kappa}\right)_x=
\int_0^1du\,V(u,x)\,\phi(u,Q),
\end{equation}
the general solution of which is  known as
\begin{equation}
\phi(x,Q)=x(1-x)\sum_n\phi_n(Q)C_n^{(3/2)}(2x-1);
\end{equation}
with Gegenbauer polynomials   $C_n^{(3/2)}$ (only even n survive) and the
$Q-$ dependence is known as
\begin{equation}
\phi_n(Q)=\phi_n(\mu)e^{A_n\kappa}=\phi_n(\mu)\left(
{\alpha_S(\mu^2)\over\alpha_S(Q^2)}\right)^{\lambda_n},
\end{equation}
where the  exponents in the expansion are ordered as
\begin{equation}
{\lambda_0}=0 ~~> {\lambda_2}=-0.62~~>~{\lambda_4}\ \ldots
\end{equation}

Calculating the integral
\begin{equation}
\int_0^1dx\,\phi(x,Q)=\phi_0(Q)\int_0^1dx\,x(1-x)={\phi_0\over6}=
\sqrt{2}f_{\pi}
\end{equation}
one can write down the beginning of the expansion:
\begin{equation}
\phi(x,Q)=6 \sqrt{2}f_{\pi}x(1-x)+(\ln Q^2)^{-0.62}\,\Phi_2\,x(1-x)[5(2x-1)^2-1]
+\ldots
\end{equation}
The pion asymptotic distribution, when $Q\rightarrow\infty$,  is then
\begin{equation}
\phi(x,Q\rightarrow\infty)\sim 6\sqrt{2}f_{\pi}x(1-x).
\end{equation}
This however does not tell us much on the realistic distribution amplitude
at accessible energies: the constants $\Phi_2,\ldots\Phi_n$ are unknown.

This is how far perturbative QCD can lead us about
the distribution amplitude $\phi$; {\it i.e.} to understand how
strong interactions build a hadron from its valence quarks. To go further, one
needs other methods, which are non perturbative by nature. Experiments can
guide us to
develop new ways since exclusive scattering data may be processed to extract
distribution amplitudes. The existing methods, like lattice calculations or QCD
sum rules, are still too primitive and rely on too many unchecked hypotheses to
be trusted. They however lead to useful rate estimates. They generally
evaluate {\it moments} of the distribution amplitude defined as:
\begin{equation}
\int_0^1dx(2x-1)^2\phi(x,\mu),\ \ldots
\end{equation}
Such a study lead Chernyak and Zhitnitsky~\cite{Che84} to propose the
distribution
\begin{equation}
\phi_{\rm cz}(x,Q^2)=6\sqrt{2}f_{\pi}x(1-x)\left\{1+[5(2x-1)^2-1]\left(
{\ln Q^2/\Lambda^2\over\ln Q_0^2/\Lambda^2}\right)^{-0.62}\right\},
\end{equation}
with $Q_0\approx500$MeV.

\subsection{Transverse Degrees of Freedom}

A study of one loop corrections~\cite{Dit81} leads to propose that
 the scale relevant for the running coupling constant $\alpha_S$
is more likely to be the exchanged gluon  virtuality $xx'Q^2$ than the
photon virtuality $Q^2$.  The whole treatment would then be correct
only when the gluon is far off mass shell, that is as far as $x$ or $x'$ do
not approach 0.  However, for intermediate transfers, it turns out that
an important part of the amplitude comes from these regions. One should
thus reexamin the whole story in the region where gluons may become soft.
In this region transverse momentum (or transverse distance) degrees of freedom
become important and invalidate the collinear approximation~\cite{Li92}.
Let us qualitatively explain the expected  modifications.

 The elastic interaction of a coloured object (a quark for instance) is
suppressed by a Sudakov form factor~\cite{Sud56} which quantifies the difficulty
 of preventing an accelerated charge  from  radiating.
Similarly the elastic interaction of a dipole of  transverse size $b$
is strongly suppressed unless $b$ approaches $Q^{-1}$~\cite{BS89}.
The approximation where transverse  degrees of freedom are
 neglected leads to consider the region
$b^2 \leq (xx'|q^2|)^{-1}$, which is an unsuppressed region
when $xx'$ is of order 1. When $xx'\rightarrow 0$, this
approximation becomes illegitimate, and one should envisage to compute
the hard scattering without freezing the transverse degrees of freedom
and use the $b$-dependence of the wave function~\cite{Li92}. One gets
(with some technically justified approximations)
\begin{equation}
T(-xx'q^2,b)\approx {2\over 3}\pi\alpha_S(t)C_F\,K_0(\sqrt{-xx'q^2}\,b),
\end{equation}
where $K_0$ is a modified Bessel function.

The interest of this improved approach is that taking radiative corrections
grouped in the wave  function into account, and analyzing
through the renormalization group the pertinent scale for the coupling
$\alpha_S$ in the above expression of $T$, one gets
\begin{equation}
t=\max(1/b,\sqrt{xx'|q^2|}).
\end{equation}
The Sudakov suppression of large transverse sizes enforces then the form factor
to receive sizeable contributions at large  transfer ($> 5\,$GeV$^2$),
only from the region where  $b$ is sufficiently small. The scale $t$ of the
perturbative approach then remains large enough in the whole relevant
integration domain.

\subsection{Experimental review}
 Precise experiments on various exclusive amplitudes
are needed to determine the distribution amplitudes. Data for the
electromagnetic form
 factor of the pion are yet too poor to even choose between the asymptotic
and the
 CZ forms. This is due to the difficulty of handling a pion target, which
is only
available through a model-dependent extraction of some quasi-real pion pole
exchange
in forward pion electroproduction.
The case for the $\pi \gamma $ transition form factor is far better. A
quite complete
 analysis~\cite{Ong,KR}
indicates that the good data obtained at $e^+ e^-$ colliders in $\gamma
\gamma $ collisions
are well described by the asymptotic distribution for $Q^2$ in the range $2
- 8 GeV^2 $, as shown
in Figure 4.

\vglue5cm
\centerline{\small Figure 4: The $\pi \gamma $ transition form factor
\cite{KR}}

\section{Other hard scattering processes. }

The results obtained above for the  electromagnetic form factor may be
generalized to other hard exclusive  processes, with an important difference
in the case of  hadron - hadron collisions~\cite{Lan,BS89}  . One thus
defines a distribution amplitude for the proton and analyzes the magnetic
form factor $G_M$ very similarly. One can then  consider  sharper
 reactions as real or virtual  Compton scattering, which still only depend on
the proton structure but where one can vary dimensionless ratio such as angles.

\subsection{The proton distribution amplitude}

As for the  pion case, the  valence nucleon wave function can be
written~\cite{Gou}
as a combination of definite tensors of colour, flavour and spinor indices
with a
(unique) proton distribution amplitude $\phi(x,y,z)$.
This distribution amplitude may be written as an expansion quite similar to
what
was derived above for the pion case but on a different basis of
polynomials, and
without the help pf weak decay processes to fix the normalization:
\begin{eqnarray}
&\phi(x_i,Q)= 120 x_1 x_2 x_3 ~\delta(x_1+x_2+x_3-1) \times \\
 & \left[\left({\alpha_S(Q^2)\over\alpha_S(Q_0^2)}\right)^{\lambda_0} A_0 +
 {{21}\over{2}}
\left({\alpha_S(Q^2)\over\alpha_S(Q_0^2)}\right)^{\lambda_1} A_1
 P_1(x_i)+
{{7}\over{2}} \left({\alpha_S(Q^2)\over\alpha_S(Q_0^2)}\right)^{\lambda_2}
 A_2 P_2(x_i) + \dots \right] \nonumber,
\label{Appell}
\end{eqnarray}

\noindent
where the slow $Q^2$ evolution comes entirely from the terms
$ \alpha_S(Q^2)^{\lambda_i}$, and the  $ \lambda_i$'s are decreasing numbers:
\begin{equation}
 \lambda_0 = {{2}\over{27}}~~<~~\lambda_1 = {{20}\over{81}}   ~~<~~\lambda_2
 = {{24}\over {81}} ~~... ,
\end{equation}
and the $P_i(x_j) $'s are Appell polynomials:
\begin{equation}
P_1(x_i)=x_1-x_3 ~~~,~~~ P_2(x_i)=1-3x_2 ,...
\end{equation}

\subsection{The proton magnetic  form factor}

One  describes the elastic interaction
of a proton and an electron
with two form factors $F_1$, which is helicity conserving, and $F_2$, which is not:
\begin{equation}
\langle p',h'|J^{\mu}(0)|p,h\rangle=e\bar{u}(p',h')\left[
F_1(Q^2)\gamma^{\mu}+i{\kappa\over2M}F_2(Q^2)\sigma^{\mu\nu}(p'-p)_{\nu}
\right]u(p,h);
\end{equation}
or their linear combinations, $G_M$ and $G_E$
called the Sachs form factors.
$h$ and $h'$ are respectively the incoming and outgoing proton helicities,
 $u$ and $\bar{u}$ their spinors and  $M$ the
proton mass. In this decomposition, $e$ is the proton charge and
 $\kappa=1.79$ is its anomalous magnetic moment.
 In the  formalism we are presenting here, only the helicity conserving form
factor $F_1$ is easily accessible. There is much discussion on the
phenomenological
analysis of experimental data, as shown on Figure 5. On the one hand, the
$Q^{-4}$
 scaling behaviour is clearly seen in the $Q^2$ range $5 - 30 GeV^2$. The
slight
decrease of $Q^4 G_M(Q^2)$ may even be understood
as a manifestation of radiative corrections on top of the counting rules.
On the other hand, the proposed understanding of the normalization of $Q^4
G_M(Q^2)$
through a much asymmetric distribution amplitude based on a QCD sum rule
analysis~\cite{Che84},
leads to many problems. Does-it mean that $F_1$ is still dominated by long
distance soft
physics~\cite{soft}? Progress on this question may come from the
computation of QCD corrections to
the present leading logarithm analysis, or from the study of color transparency
 observables.

\vglue5cm
{\centerline {\small Figure 5: $Q^2$ evolution of the proton magnetic form
 factor~\cite{Sil93}}}

\subsection{Compton scattering}

The  perturbative part of the analysis of real~\cite{Far88,Kro91}
 Compton scattering consists in evaluating the 336
topologically distinct diagrams obtained when coupling two photons to the
three valence quarks of the proton, two gluons being exchanged. Moreover,
there are 42 diagrams with a three-gluon coupling but it turns out that
their color
factor vanishes.

There are few existing data for real Compton scattering on the proton
with $-t>1GeV^2$ \cite{Shupe}. The higher energy and transfer data seem to
approach
the dimensional scaling law (in $s^{-6}$ ) but more data at higher energies
and various
angles around $90^\circ _{CM}$ are clearly needed before one can use them
to extract
the proton distribution amplitude. To do this, one should  write the cross
section
 as a sum of terms
\begin{equation}
A_i T_H^{ij}(\theta) A_j
\end{equation}
where the decomposition of the distribution amplitude on
 the Appell polynomials (Eq.~(\ref{Appell})) has been used and where
 $T_H^{ij}$ are  integrals over   $x$ and $y$ variables of the product of
the hard amplitude at a given scattering angle
 $\theta$   by the two Appell polynomials
$A_i(x)$ and $A_j(y)$. The $T_H^{ij}$ are ugly long expressions but they
can be numerically handled.

Determining the proton distribution amplitude from experimental data
boils down then to the extraction by a maximum of likelyhood method  of
the  $A_i$ parameters, amputating the series of Eq.~(\ref{Appell}) to its
first $n$  terms, by taking advantage of the (admittedly mild) $Q^2$
dependence of the  $A_i$ parameters which indicates that they decrease
quicker for large $Q^2$ at larger $n$. One should of course verify
afterwards that
including the term $n+1$
does not drastically modify the conclusion. One can then
explore other reactions, virtual Compton scattering for instance,
which must be well described by the same series of $A_i$'s.

\

\subsection{Virtual Compton Scattering}
An intense electron accelerator such as ELFE would enable us to go further
and study virtual
Compton scattering in the large angle region. At lowest order in  $\alpha
\sim {{1} \over {137}}$, Virtual Compton
Scattering  (VCS) is described as the  coherent sum of two amplitudes
 namely the Bethe Heitler (BH) process
where the  final photon is radiated from the electron and the genuine
VCS process where it comes from the proton.
The perturbative calculation ~\cite{Far91} of the VCS amplitude is not much
harder than the real case. As the BH amplitude is calculable from the elastic
 form factors $G_{Mp}(Q^2)$ and $G_{Ep}(Q^2)$,
 its interference with the VCS amplitude is an  interesting source of
information, different from what real
Compton scattering yields. The VCS amplitude depends on three invariants;
 one usually chooses $ Q^2,  s, t $ or $s, Q^2/s, \theta_{CM} $.

The forward region is also interesting and is currently subject to intense
theoretical
investigation; the counting rates are expected to be larger but the Bethe
Heitler
process constitutes there a large background.

\subsection{Other processes }

Photo- and electro-production of mesons at large angle will allow to
probe  distribution amplitudes of $\pi$ and $\rho$ mesons in the same way.
The production of the $ K \Lambda $  final state  selects a few
 hard scattering diagrams. The analysis of these reactions is still to be
done if one excepts some works done in the simplifying framework of the
diquark model\footnote{The physical idea behind the diquark model is not to
consider a new field theory
where diquarks would be fundamental. It is mostly to phenomenologically
take into
account of quark correlations inside baryons which seem to be indicated by
 experimental data, and thus modelize the proton wave function as a
quark-diquark composite,
while adopting the general framework indicated by QCD for exclusive
reactions, that is
assume factorization holds and calculate the hard subprocess with some new
Feynman rules
for diquark-gluon or diquark-photon vertices.
It is not obvious at all, to say the least, that such a procedure is
consistent. One
 thus should take calculations in this type of approach as useful exercises
which may
turn out to explain part of the physical reality is indeed quark-quark
correlations
are playing a major role in some exclusive processes at medium transfers.
A number of calculations have been performed within the diquark picture and
we cannot
discuss all of them here. It turns out that treating baryons as quark-diquark
objects render computations much easier, in particular because the number
of Feynman
 diagrams is quite reduced. This is particularly true if only scalar
diquarks are
considered. We refer the interested reader to the existing
litterature~\cite{Diq}.}.

\section{Color transparency}
\subsection{The physical idea}
Hard exclusive scattering ( with a typical large $Q^2$ scale)  selects a very
  special quark configuration: the minimal valence state where
all quarks are  close together,   a  small size  color
neutral  configuration  sometimes  referred  to  as a {\em mini hadron}.

Such  a color  singlet system  cannot emit  or absorb  soft gluons
which carry energy or momentum smaller than $Q$.  This is because  gluon
radiation --- like photon radiation in QED --- is a coherent process and
there is thus destructive interference between gluon emission amplitudes
by quarks  with ``opposite''  color.   Even without  knowing exactly how
exchanges  of   soft  gluons   and  other   constituents  create  strong
interactions, we  know that  these interactions  must be  turned off for
small color singlet objects. This is color transparency~\cite{MU82,jpr}.

An exclusive hard reaction will thus probe the structure of a {\em  mini
hadron}, i.e. the short distance part of a minimal Fock state  component
in the hadron  wave function.   This is of  primordial interest for  the
understanding  of  the  difficult  physics  of  confinement.    First,
selecting the simplest Fock state amounts to the study of the  confining
forces in  a colorless  object which is quite reminiscent of the
``quenched  approximation'' much used in lattice QCD simulations, where
quark-antiquark pair creation from  the vacuum is forbidden.   Secondly,
letting the mini-state  travel through different nuclei
of various  sizes allows  an indirect  but unique  way to  test how  the
squeezed mini-state  interacts with hadrons, opening a new chapter of
strong interaction physics.

To the extent that the electromagnetic form factors are understood as  a
function of $Q^2$,
\begin{equation}
e+A \rightarrow e+(A-1)+p
\end{equation}
experiments will measure
the color screening properties of QCD.   The quantity to be measured  is
the transparency ratio $T_r$ which is defined as:
\begin{equation}
T_r = \frac{\sigma_{\rm Nucleus}}{Z \sigma_{\rm Nucleon}}
\end{equation}

At asymptotically large values  of $Q^2$, $T_r$ should approach unity.
The  approach to the scaling behavior as well as  the value of $T_r$ as a
function  of the  scaling  variable  determine  the  evolution  from  the
pointlike configuration to the  complete hadron. We will not present
here the many ideas which have recently emerged in this new field~\cite{jpr}.

\subsection{An instructive calculation}

Let us now present a somewhat academic but instructive calculation of the
high energy
limit of a forward amplitude in perturbative field theory~\cite{ct}.
Although such a
computation would be more reliable in QED, let us pretend it is going to
describe the
strong interactions of hadrons. The idea is to use the optical theorem to
relate the
total cross section to the imaginary part of the forward amplitude.

The first step is to calculate in the region $-t << s \sim -u $ the
scattering amplitude
of quark-quark scattering through one gluon exchange(Fig.6a); the result is
(omitting a
 factor specifying helicity conservation):
\begin{equation}
{\cal M} = 2 C g^2 s {1 \over {t-\lambda^2}}
\end{equation}
which may be rewritten, but this is more a mathematical trick than a
physical idea
 at this stage, as

\begin{equation}
{\cal M} = -2 C  s \int d^2b e^{iq.b} {{g^2 K_0 (\lambda b)}\over {2\pi}}
\end{equation}
where a $2-$dimensional Fourier transform in transverse space has been
performed, writing
$t = - q^2$ with $q$ a $2-$dimensional transverse vector.
 $C$ is a color factor, $g$ the quark gluon coupling and $\lambda$ an
effective gluon
 mass.
Let us now consider  two gluon exchange processes. From the 18 diagrams
which may be drawn,
14 are vertex or self energy corrections to the one-gluon diagrams, and as
such are
real. Two of the four remaining diagrams dominate at small $(-t)$ and are
drawn on Fig.6b. Denoting as $q$ the $4-$vector of the first gluon emitted
by the
bottom fermion line, one finds that the two diagrams have identical boson
propagators and
bottom fermion expression. From the upper fermion line, one gets adding the
two diagrams,
after some simple algebra:
\begin{equation}
{\cal A} \propto {1 \over {q^- +i\epsilon}} - {1 \over {q^- -i\epsilon}}
\end{equation}
leading to a $\delta(q^-)$ constraint. Moreover the bottom fermion line
leads to a
$\delta(q^+)$ constraint, so that the $d^4q$ integration boils down to a
$2-$dimensional
transverse space integration. The resulting amplitude is dominantly imaginary:

\begin{equation}
{\cal M} = i C  s \int {{d^2q_T} \over {(2\pi)^2}} {1 \over {(q^2_T+\lambda^2)
((q_T-\Delta_T)^2+\lambda^2)}}
\end{equation}
which may be rewritten as
\begin{equation}
{\cal M} = i C  s \int d^2b e^{i\Delta_T.b} ({{g^2 K_0 (\lambda b)}\over
{2\pi}})^2
\end{equation}
where $\Delta_T$ is the (small) $2-$dimensional transfer between the
initial and
final fermion.

The important features of this result is the $2-$dimensional nature of the
integral
 and the squared factor  reminiscent of the
eikonal nature of forward amplitude in the high energy limit.

\vglue5cm
\centerline{\small Figure 6: quark-quark and meson-meson scattering}

\vspace{.2in}
Let us now go to the almost physical case of meson-meson scattering, that
is consider  a
color singlet formed by a quark -antiquark pair scattering on another pair
(Fig. 6c). A
straightforward computation shows that the color factors are the same for
all $2-$gluon
exchange graphs connecting the upper hadron to the lower one, but that an
extra minus sign
is attached for an antiquark line. The result is then
\begin{eqnarray}
&{\cal M} = i C  s \int d^2b e^{i\Delta_T.b}\int d^2r_1d^2r_2
\psi^*(r_1)\psi^*(r_2) \\
&\{V(x_1-x_3)-V(x_2-x_3)-V(x_1-x_4)+V(x_2-x_4)\}^2 \psi(r_1)\psi(r_2)\nonumber
\end{eqnarray}
with
\begin{equation}
V(x) = {1 \over {(2\pi)^2}} \int {{d^2k e^{ik.x}} \over {k^2+ \lambda^2}}
\end{equation}
Suppose now that $r_1 = x_1-x_2$ is small (a mini-hadron); then one may
approximate
$(V(x_1-x_3)-V(x_2-x_3) \sim r_1. \nabla (x_1-x_3) $ and similarly for
$V(x_1-x_4)-V(x_2-x_4)$, and get:
\begin{equation}
{\cal M} \propto r_1^2 .
\end{equation}
leading to a total cross section for the mini hadron scattering on a hadron:
\begin{equation}
\sigma \sim {Im {\cal M}\over  s} \propto r_1^2 .
\label{ct}
\end{equation}
This is color transparency. Note that if the $q \bar q$ was in a color
octet state,
the color factor would still be factorized but the resulting amplitude
would contain
an additionnal term
\begin{equation}
{\cal M} \propto  \dots + \{(V(x_1-x_3)-V(x_1-x_4)\}.\{V(x_2-x_3)-V(x_2-x_4)\}
\end{equation}
which does not vanish in the limit $x_1 \rightarrow x_2$. The color singlet
nature of the
quark-antiquark pair is thus essential for Eq.\ref{ct} to be valid.

\subsection { Present Data and future prospects}

\hspace {\parindent}
Experimental data~\cite{ctexp} on color transparency are very scarce and,
although they
have been discussed at length~\cite{jpr}, not yet conclusive.
 Color transparency is indeed just an
emerging field of study and one should devote much attention to get
more information on this physics in the near future.

A second round of proton experiments at Brookhaven has been approved and
 a new detector named Eva ~\cite{BNL850} with much higher acceptance
 has been taking
data for about one year. Along with other improvements and increased
beam type, this should increase the amount of data taken by a quite large factor
allowing a wider energy range and an analysis at different
scattering angles. It would also be very interesting to analyze meson-nucleus
 scattering and their spin properties in similar conditions~\cite{gpr}.

The Hermes detector ~\cite{HERMES} at HERA is beginning operation. It should
enable a confirmation of FNAL data on $\rho$ meson diffractive production
at moderate $Q^2$ values and quite smaller values of energies
$10 \leq \nu \leq 22\,$GeV.

The use of nuclear targets to test  color transparency and use nuclear filtering
is one of the major goals of the $15$--$30\,$GeV continuous electron beam ELFE
project~\cite{ELFE}. The $(e,e'p)$ reaction
should in particular be studied in a wide range of $Q^2$ up to $21\,$GeV$^2$,
thus allowing to connect to SLAC data (and better resolution but similar low
$Q^2$ data from CEBAF) and hopefully clearly establish this phenomenon
in the simplest occurence.

 Using light nuclei such as deuteron or helium puts
the emphasis on the reduction of final state interaction effects which are
quite well
under control in this case. The measure of the recoil momentum spectrum
of a spectator neutron when a proton has experienced a hard scattering, for
instance,
 is a direct test of the decrease of the subsequent mini-proton neutron elastic
scattering cross section.

The measurement of the transparency ratio for photo- and
electroproduction of heavy vector mesons, in particular of $\psi$ and
 $\psi'$ will open  a new regim where the mass of the quark enters as
a new scale controlling the formation length of the produced meson.

\section{Acknowledgements}
I acknowledge useful discussions with many colleagues and I specially thank
 Thierry Gousset and John P. Ralston for much work in collaboration on this
subject.


\begin{thebibliography}{99}

\bibitem{ELFE}  {\it The ELFE Project} Conference Proceedings,
 Vol.44, Italian Physical Society, Bologna, Italy (1993) edited by J.
Arvieux and E. DeSanctis;
J. Arvieux and B. Pire, Progress in Particle and Nuclear Physics, {\bf 30},
 299 (1995).

\bibitem{Bro73}
S.J. Brodsky and G.R. Farrar, Phys. Rev. Lett. {\bf 31}, 1153 (1973);
V.A. Matveev, R.M. Muradyan and A.V. Tavkhelidze, Lett. Nuovo
Cimento {\bf 7}, 719 (1973).

\bibitem{Lan}
 P.V. Landshoff, Phys. Rev. {\bf  D10}, 1024 (1974).

\bibitem{Mue81}
 A.H. Mueller, Phys. Rep. {\bf 73}, 237 (1981).

\bibitem{BS89}
J. Botts and G. Sterman, Nucl. Phys. {\bf B325}, 62 (1989).

\bibitem{Far79}
G.R. Farrar and D. Jackson, Phys. Rev. Lett. {\bf 43}, 246 (1979);
S.J. Brodsky and G.P. Lepage, Phys. Lett. {\bf 87B}, 359 (1979);
A.V. Efremov and A.V. Radyushkin, Phys. Lett. {\bf 94B}, 245 (1980);
V.L. Chernyak, V.G. Serbo and A. R. Zhitnitsky,
Yad. Fiz. {\bf 31}, 1069 (1980);
A. Duncan and A.H. Mueller, Phys. Rev. {\bf  D21}, 1636 (1980).

\bibitem{Sal51}
E.E. Salpeter and H.A. Bethe, Physical Review {\bf 84}, 1232 (1951).

\bibitem{Bro}
S.J. Brodsky and D.G. Robertson, {\sl Proceedings of the ELFE
 Summer School on Confinement Physic} edited by S.D. Bass and P.A.M.
Guichon (Ed. Fronti\`eres,
Gif, 1996).

\bibitem{Gou}
T. Gousset, ``Hadron wave function in hard exclusive scattering'',{\sl
Proceedings of the ELFE
 Summer School on Confinement Physic},edited by S.D. Bass and P.A.M.
Guichon (Ed. Fronti\`eres,
Gif, 1996).

\bibitem{Field}
R.D. Field, {\sl Applications of Perturbative QCD}
(Addison-Wesley, Redwood, 1989).

\bibitem{Che84}
V.L. Chernyak and A.R. Zhitnitsky, Phys. Rep. {\bf 112}, 173 (1984).

\bibitem{Dit81}
F.-M. Dittes and A.V. Radyushkin, Sov. J. Nucl. Phys. {\bf 34}, 293 (1981).

\bibitem{Li92}
H.-N. Li and G. Sterman, Nucl. Phys. {\bf B381}, 129 (1992).

\bibitem{Sud56}
V.V. Sudakov, Sov. Phys. JETP {\bf 3}, 65 (1956).

\bibitem{Ong}
S. Ong, Phys. Rev. {\bf  D52}, 3111  (1995).

\bibitem{KR}
 R. Jakob, P. Kroll and M. Raulfs, J.Phys. {\bf G22}, 45 (1996); P. Kroll
and M. Raulfs,
Wuppertal preprint hep-ph/9605264.

\bibitem{soft}
N. Isgur and C.H. Llewellyn-Smith, Nucl. Phys. {\bf B317}, 526 (1989); J.
Bolz{\it et\ al.}
Z. Phys. {\bf C66} 267 (1995).

\bibitem{Sil93}
A.F. Sill {\it et\ al.}, Phys. Rev. {\bf  D48}, 29 (1993).

\bibitem{Far88}
G.R. Farrar and E. Maina, Phys  Lett {\bf B206} , 120 (1988);
G.R. Farrar and H. Zhang, Phys Rev Lett {\bf 65}, 1721 (1990);
G.R. Farrar and H. Zhang, Phys Rev {\bf D41}, 3348 (1990).

\bibitem{Kro91}
A.S. Kronfeld and B. Ni\v zi\'c, Phys Rev {\bf D44}, 3445 (1991); erratum
{\bf D46}
 2272 (1992).
%

\bibitem{Shupe}
M.A. Shupe {\it et\ al.}, Phys. Rev. {\bf  D19}, 1929 (1979).

\bibitem{Far91}
G.R. Farrar, K. Huleihel and H. Zhang, Nucl. Phys. {\bf B349}, 655 (1991);
P. Kroll, M. Schurmann and P. Guichon,  Nucl. Phys. {\bf A598}, 435 (1996).


\bibitem{Diq}
 P.Kroll, M.Schurmann and W. Schweiger, {\it Proceedings of Quark Cluster
 Dynamics}, (Bad Honnef, Germany 1992), edited by K. Goeke (Springer-Verlag,
1992) p 179; M. Anselmino {\it et al.}, Rev. Mod. Phys.66, 195 (1993).
{\it Proceedings of the Workshop on Diquarks}, M. Anselmino and E.
Predazzi, ed. World
 Scientific(1989); {\it Proceedings of the Workshop on Diquarks II}, M.
Anselmino and E. Predazzi, ed. World
 Scientific(1994).

\bibitem{MU82} A. Mueller, {\it Proceedings of the Seventeenth Recontre de
Moriond on Elementary Particle Physics}, (Les Arcs, France 1982) edited by
J. Tran Thanh Van (Editions Frontieres, Gif-sur-Yvette, France 1982), p. 13;
S. J. Brodsky, in {\it Proceedings of the Thirteenth International Symposium
on Multiparticle Dynamics}, edited by W. Kittel, W. Metzger, and A. Stergiou
(World Scientific, Singapore, 1982) p. 963.
%
\bibitem{jpr}
 P. Jain, B. Pire and J.P. Ralston,  Physics Reports {\bf 271}, 67 (1996);
see also L. Frankfurt, G.A. Miller, and M. Strikman, Comments Nucl. Part. Phys.
{\bf 21}, 1 (1992).

\bibitem{ct}
T.P. Cheng and T.T. Wu, {\it Expanding protons}, MIT press, Cambridge
(1987); J.F. Gunion and D.E. Soper ,Phys Rev {\bf D15}, 2617 (1977); J.
Dolejsi and J. Hufner,
Z. Phys  {\bf C54}, 489 (1992).
%

\bibitem{ctexp}
 A.S. Carroll {\it et\ al.}, Phys. Rev. Lett. {\bf 61}, 1698 (1988);
 T.A. Armstrong {\it et al.}, Phys. Rev. Lett. {\bf 70},
1212 (1993);
M.R. Adams {\it et\ al.}, Phys. Rev. Lett. {\bf 74} , 1525 (1995).

\bibitem{BNL850} {\it BNL--Experiment 850}; spokesmen A. Carroll and S.
Heppelmann.

\bibitem{gpr} T. Gousset, B. Pire and J.P. Ralston,  Phys.
Rev.{\bf D 53}, 1202 (1996).


\bibitem{HERMES}
 {\it Hermes proposal}, Report DESY--PRC 90/01.






\end{thebibliography}
\end{document}